\documentclass[epj]{svjour}
\usepackage{psfig}
\usepackage{times}
\begin{document}
\title{On the current status of OZI violation in ${\pi}N$ and $pp$ 
reactions\thanks{Supported by  Forschungszentrum  J\"ulich}}
\author{A. Sibirtsev and  W. Cassing}
\institute{Institut f\"ur Theoretische Physik, Universit\"at Giessen \\
D-35392 Giessen, Germany}
\date{Received: date / Revised version: date}

\abstract{
The available data on $\omega$ and $\phi$ production from 
$\pi N$ and $pp$ collisions are reanalyzed with respect to an 
OZI rule violation on the basis of transition matrix elements. 
The data are found to be compatible with a constant ratio 
$R$, which however, deviates substantially from the SU(3) 
prediction based on the present knowledge of the $\phi{-}\omega$ 
mixing angle.
}
\authorrunning{A. Sibirtsev and  W. Cassing }
\titlerunning{On the current status of OZI violation in 
${\pi}N$ and $pp$  reactions}

\maketitle

\section{Introduction}
Assuming the ideal  SU(3) octet-singlet mixing Okubo, Zweig and 
Iizuka proposed~\cite{Okubo,Zweig,Iizuka} that the production of 
a $\phi$-meson from an initial non-strange state is strongly 
suppressed in comparison to $\omega$-meson production. 
Indeed, because of SU(3) breaking the octet and singlet states
are mixed and for an ideal mixing angle $\theta_V{=}35.3^0$ the 
$\phi$-meson is a pure  $s\overline{s}$ state.  In case of  
$\phi$ production  from ${\pi}N$, $NN$ or $N\bar{N}$ reactions 
the OZI rule states that the contribution from  the diagram with a
$s\overline{s}$ pair disconnected from
the initial $u,d,\bar{u},\bar{d}$ should ideally vanish. 
The experimental deviation from the ideal mixing angle
$\Delta\theta_V$=3.7$^0$~\cite{PDG} can be used~\cite{Lipkin}  to 
estimate the ratio $R(\phi{/}\omega){\approx}4.2{\times}10^{-3}$ 
of the cross sections with a $\phi$ and 
$\omega$ in the final state. This deviation of the experimental 
ratio $R$ from zero is denoted as OZI rule violation. A large ratio $R$
might indicate an intrinsic $s\overline{s}$ content of the nucleon since
in that case the $\phi$-meson production is due to
a direct strangeness transfer from the initial to the 
final state and thus OZI allowed.

The OZI violation problem has lead to a large experimental activity
involving different hadronic reactions. Here we perform
a systematical data analysis for ${\pi}N$ and $pp$ reactions
and discuss their theoretical interpretation in 
context with the most recent data point
from the DISTO Collaboration~\cite{Balestra}. 

\section{$\omega$ and $\phi$ production in ${\pi}N$ reactions}
Without involving any theoretical assumption about the production 
mechanism the data~\cite{LB} on the total ${\pi}N{\to}{\omega}N$  
and ${\pi}N{\to}{\phi}N$  cross sections may be analyzed
in terms of the corresponding transition amplitudes. The amplitude for 
a two-body reaction with  stable particles in the final state
is related to the total cross section $\sigma$ as~\cite{Feynman}
\begin{equation}
|M_V| = 4 \left[\pi \sigma s \right]^{1/2} \left[
\frac{\lambda (s,m_N^2,m_\pi^2)}{\lambda (s,m_N^2,m_V^2)}
\right]^ {1/4},
\label{two}
\end{equation}
where $\lambda(x,y,z)=(x-y-z)^2-4yz$, while $m_N$, $m_\pi$, $m_V$
denote the nucleon, pion and vector meson masses, respectively,
and $s$ is the squared invariant collision energy. Moreover, we
compare the transition amplitudes for $\omega$ and $\phi$
production at the same  excess energy 
$\epsilon{=}\sqrt{s}{-}m_N{-}m_V$. As was discussed in
Ref.~\cite{Hanhart}, Eq.~(\ref{two})
can be used for the evaluation of the amplitudes for 
the production of unstable ($\omega$ and $\phi$) mesons
at excess energies $\epsilon {>}\Gamma_V$, where $\Gamma_V$ 
denotes the width of the vector meson spectral function 
due to its vacuum decay. 

Furthermore, due to the experimental set up the 
$\pi^-p{\to}{\omega}n$ data from Ref.~\cite{Karami} should not 
be considered as total cross sections, but as differential 
cross sections $\sigma_{dif}$ integrated over a given 
range of the final neutron momentum~\cite{Hanhart}. Indeed,  
the $\pi^-p{\to}{\omega}n$
cross sections  given in Ref.~\cite{Karami} for
different intervals $[q_{min},q_{max}]$ of neutron
momenta in the center-of-mass system  can be
related to the transition amplitude $M_V$ as
\begin{eqnarray}
\sigma_{dif}=\intop_{q_{min}}^{q_{max}}
\frac{|M_V|^2}{4 \pi^2 \, \lambda^{1/2}(s,m_p^2,m_\pi^2)} \,
\frac{q^2}{\sqrt{q^2+m_n^2}}
\nonumber \\
\frac{\Gamma_V m_V}{(s-2\sqrt{s(q^2+m_n^2)}+m_n^2-m_V^2)^2
-\Gamma_V^2m_V^2} \ dq,
\label{chris}
\end{eqnarray}
where $m_p$ and $m_n$ are the proton and neutron masses, 
respectively, and $s$ is given as a function of $q$.
Eq.~(\ref{chris}) agrees with that in Ref.~\cite{Hanhart}
in the non-relativistic limit. Furthermore, in the calculations
 we use the set of the neutron momentum intervals $[q_{min}, q_{max}]$ 
as in Ref.~\cite{Karami}.

\begin{figure}[h]
\phantom{aa}\vspace{-0.8cm}\hspace{-6mm}
\psfig{file=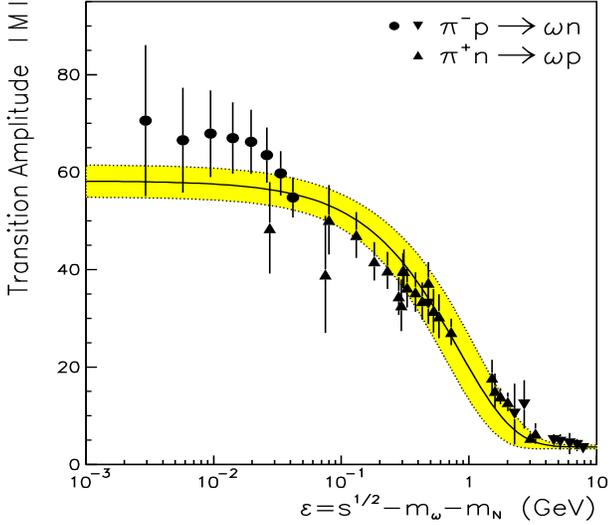,width=9.7cm,height=8cm}
\caption[]{Data on the ${\pi}N{\to}{\omega}N$  transition amplitude 
$|M|$ as a function of the excess energy
$\epsilon$.  The triangles show the data from~\cite{LB} evaluated by
Eq.~(\ref{two}) while the full dots show the data from 
Ref.~\cite{Karami} evaluated by Eq.~(\ref{chris}). 
The solid line displays the approximation~(\protect\ref{fit})
while the dashed area illustrates the uncertainty of the fit.}
\label{ozi1b}
\end{figure}

\begin{figure}[h]
\phantom{aa}\vspace{-1.4cm}\hspace{-6mm}
\psfig{file=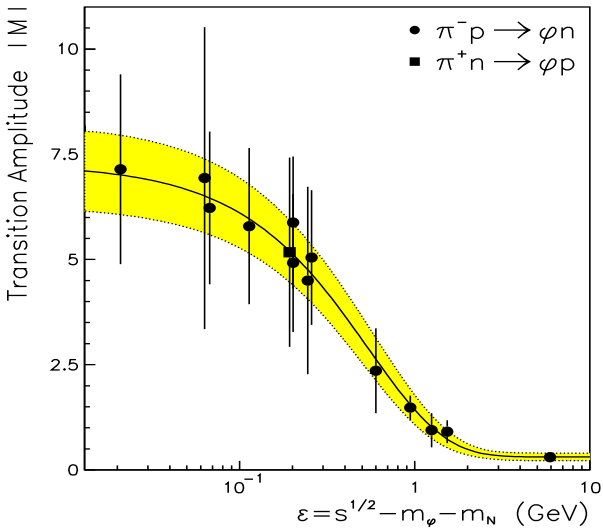,width=9.6cm,height=8cm}
\caption[]{Data on the ${\pi}N{\to}{\phi}N$  
transition amplitude $|M|$ as a function of the excess energy
$\epsilon$. The solid line shows the approximation~(\protect\ref{fit})
while the dashed area indicates the uncertainty of the fit.}
\label{ozi1c}
\end{figure}

Figs. \ref{ozi1b},\ref{ozi1c} show the transition amplitudes for 
the ${\pi}N{\to}{\omega}N$  and ${\pi}N{\to}{\phi}N$ 
reactions evaluated from the experimental data \cite{LB,Karami}. 
Note, that the $\pi^-p{\to}{\omega}n$ transition amplitude 
evaluated from the data of Ref.~\cite{Karami} (full 
dots at small $\epsilon$) by Eq.~(\ref{chris})
does not depend on energy within the errorbars and agrees well with that
extracted from the other data~\cite{LB}.

Since the data are not available for a comparison at exactly the 
same excess energies we fit the transition amplitudes by the
function
\begin{equation}
|M_V| = M_0 + M_1\exp{(-\gamma \epsilon)}
\label{fit}
\end{equation}
with the parameters given in Table \ref{tab1}. The solid lines in 
Figs.~\ref{ozi1b}, \ref{ozi1c} show the 
approximation~(\ref{fit}) while the 
dashed areas indicate the uncertainty of the parameterization. 
Note, that the approximation is compatible 
with an almost constant transition amplitude for 
$\epsilon{<}$ 100~MeV and reasonably reproduces the experimental 
results up to  $\epsilon{=}$10~GeV.

\begin{table}
\caption{\label{tab1} The parameters of the 
approximation~(\protect\ref{fit}).}
\center{
\begin{tabular}{|c|c|c|c|}
\hline\noalign{\smallskip}
Reaction & $M_0$ & $M_1$ & $\gamma$  \\
\noalign{\smallskip}\hline\noalign{\smallskip}
${\pi}N{\to}{\omega}N$ & 3.6 & 54.6 & 1.21 \\
${\pi}N{\to}{\phi}N$ & 0.31 & 6.96 & 1.83 \\
$pp{\to}pp\omega$ & - & 37.7 & 0.27 \\
\noalign{\smallskip}\hline
\end{tabular}}
\end{table}

\begin{figure}[h]
\phantom{aa}\vspace{-0.8cm}
\psfig{file=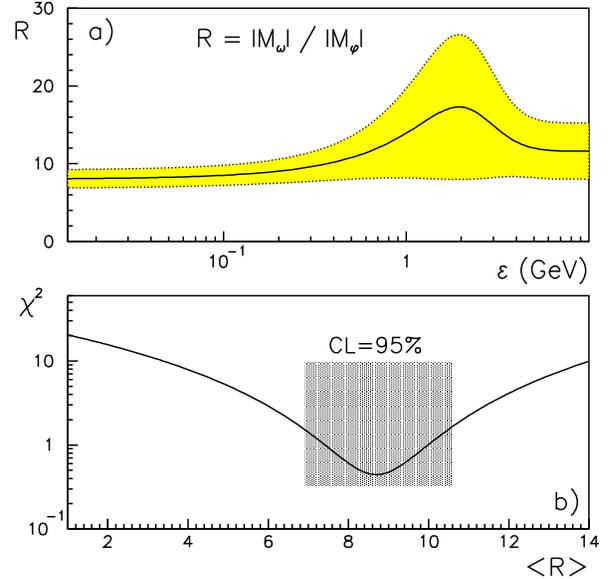,width=9cm,height=9cm}
\phantom{aa}\vspace{-0.6cm}
\caption[]{a) The ratio $R$ of the ${\pi}N{\to}{\omega}N$ and
${\pi}N{\to}{\phi}N$ transition amplitudes (solid line) and
related uncertainty ${\Delta}R$ (dashed area) as a function of the
excess energy $\epsilon$. b) The reduced $\chi^2$
for the approximation of the ratio $R$ by a constant value
${<}R{>}$ (solid line) and the confidence interval (dashed area)
for a confidence level of 95\%.}
\label{ozi1d}
\end{figure}

The resulting  ratio of the ${\pi}N{\to}{\omega}N$ to 
${\pi}N{\to}{\phi}N$ transition amplitudes is shown
in Fig.~\ref{ozi1d}a) by the solid line  as a function of 
the excess  energy $\epsilon$. It is important to note that the ratio 
$R{=}|M_\omega|{/}|M_\phi|$ is almost constant within the
given uncertainties up to  $\epsilon{=}$10~GeV, where the data 
are available.

Since the $\omega{/}\phi$ ratio is always discussed as a
constant,  that is compared to the SU(3) predictions, we
calculate the average value of ${<}R{>}$ in the range
$0{<}\epsilon{<}10$~GeV. Fig. \ref{ozi1d}b)  shows the reduced
$\chi^2$ as a function of the constant ${<}R{>}$, which
approaches a minimum at 
\begin{equation}
<R>= \frac{|M_{\pi N\to \omega N}|}{|M_{\pi N\to \phi N}|} =
8.7 \pm 1.8.
\end{equation}
with the dispersion given for a 95\% confidence level.

\begin{figure}[h]
\phantom{aa}\vspace{-7.2mm}
\psfig{file=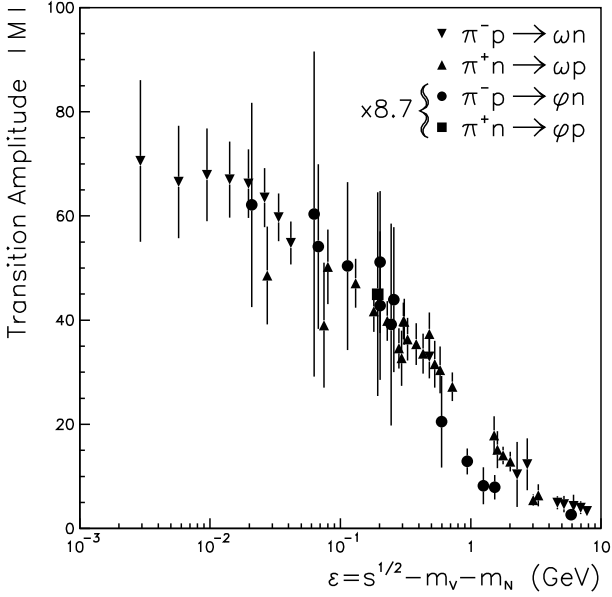,width=9cm,height=9cm}
\caption[]{Experimental 
results~\protect\cite{LB,Karami} for the 
${\pi}N{\to}{\omega}N$ (triangles) and ${\pi}N{\to}{\phi}N$ 
(circles and squares) transition amplitude $|M|$ as a function 
of the excess energy $\epsilon$, where the ${\pi}N{\to}{\phi}N$ 
amplitude is multiplied by a factor of ${<}R{>}$=8.7.}
\label{ozi1a}
\vspace{-5mm}
\end{figure}

Furthermore, a visual  way to control our estimate for  
${<}R{>}$ is to compare the experimental data directly 
by multiplying  the ${\pi}N{\to}{\phi}N$ amplitude by the factor  
${<}R{>}$ as shown in Fig.~\ref{ozi1a}. We note that four 
experimental points for the $\pi^-p{\to}{\phi}n$
reaction around $\epsilon{=}$1~GeV deviate by a factor of $\simeq$1.8
from the hypothesis applied. New experimental data with 
high accuracy are obviously necessary  for a final conclusion 
about the ratio of the ${\pi}N{\to}{\omega}N$ and 
${\pi}N{\to}{\phi}N$ reaction  amplitudes.

\section{$\omega$ and $\phi$ production in $pp$ reactions}

In our normalization the $pp{\to}ppM$ 
total cross section for the production of an {\it unstable} meson 
with total width $\Gamma$ is given as
\begin{eqnarray}
\sigma&=&\frac{1}{2^8 \ \pi^3 \ s \ \lambda^{1/2}(s,m_N^2,m_N^2)}
\hspace{-3mm}
\intop^{\sqrt{s}-2m_N}_{m_{min}} \hspace{-2mm}\frac{1}{2\pi} \
\frac{\Gamma \ dx}{(x-m_V)^2+\Gamma^2/4} \nonumber \\
&\times& \intop_{4m_N^2}^{(\sqrt{s}-x)^2} |M|^2 \
\lambda^{1/2}(s,y,x^2) \ \lambda^{1/2}(y,m_N^2,m_N^2) \nonumber \\
&\times& C^2( q=0.5\sqrt{y-4m_N^2}) \ \frac{dy}{y} ,
\label{sigma}
\end{eqnarray}
where $m_{min}$ is the minimal mass of the unstable particle
and $C(q)$ describes the final state interaction (FSI) between 
the nucleons~\cite{Watson,Migdal,GellMann,Taylor}.

\begin{figure}[h]
\phantom{aa}\vspace{-0.6cm}\hspace{-7mm}
\psfig{file=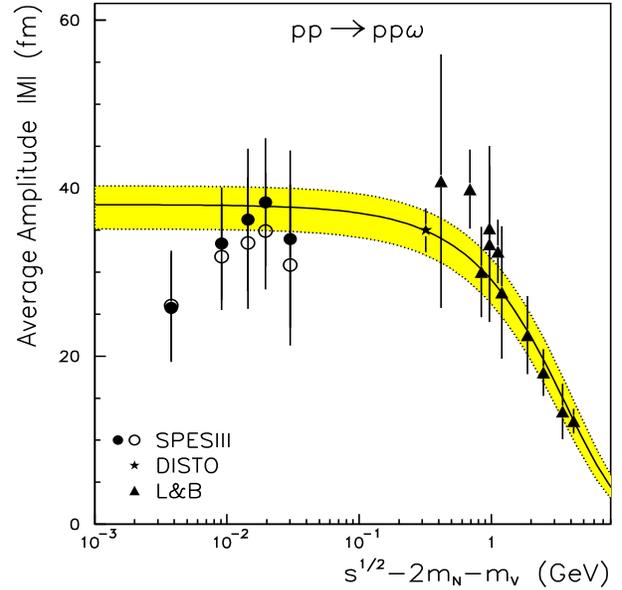,width=9.8cm,height=9.cm}
\caption[]{
The average amplitude $|M|$ for the $pp{\to}pp\omega$ reaction
as a function of the excess energy $\epsilon$. The circles
show the SPES-III~\protect\cite{Hibou} data evaluated with the
FSI model from Ref.\protect\cite{DKT} (open circles) and 
from  Ref.\protect\cite{SC1} (full circles). The triangles
indicate the data from Ref.~\protect\cite{LB}; the star is our
extrapolation for the DISTO experiment. The solid line shows the 
parameterization~(\ref{fit}) while the dashed area indicates
the related uncertainty. } 
\label{ozi2}
\end{figure}

Fig. \ref{ozi2} shows the average production
amplitude for the $pp \to pp\omega$ reaction evaluated 
by Eq. (\ref{sigma}) from the data~\cite{LB,Hibou} using the 
FSI models from Refs.\cite{DKT,SC1}\footnote{A comparison 
between the different models for the final state interaction is 
presented in Refs.\cite{SC1,SC2}}.
We note that the uncertainty in the evaluation of the
$pp{\to}pp\omega$  production amplitude due to the different
models of the FSI corrections is substantially smaller than
the dispersion of the experimental results. 

The $pp{\to}pp\omega$ reaction amplitude evaluated from the
data~\cite{LB,Hibou} is approximated by the function~(\ref{fit})
with parameters given in Tab.~\ref{tab1} and
is shown in Fig.\ref{ozi2} by the solid line. The dashed area in 
Fig.\ref{ozi2} indicates again the uncertainty of the approximation
which was calculated with the error correlation matrix. 
 
Recently the DISTO Collaboration reported an experimental 
result~\cite{Balestra} on the ratio of
the $pp{\to}pp\phi$ and $pp{\to}pp\omega$ total cross section 
at a beam energy of 2.85~GeV. For the further analysis we need
the $\phi$-meson production cross section explicitly, which can be 
obtained by normalization to the available data on 
$\omega$-meson production~\cite{LB,Hibou}. Our extrapolation
for the $pp{\to}pp\omega$ production amplitude at 2.85~GeV 
is shown in Fig.\ref{ozi2} by the star and provides
\begin{eqnarray}
\sigma (pp{\to}pp\omega)& = & 45 \pm 7 \ \ {\mu}b , \nonumber \\
\sigma (pp{\to}pp\phi) & = & 0.17^{+0.07}_{-0.06} \ \ {\mu}b .
\end{eqnarray}

\begin{figure}[h]
\phantom{aa}\vspace{-0.7cm}
\psfig{file=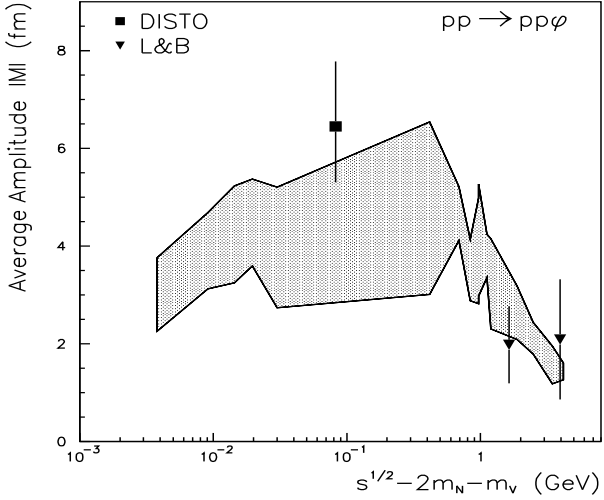,width=9cm,height=7.6cm}
\caption[]{The average amplitude $|M|$ for the $pp{\to}pp\phi$ reaction
as a function of the excess energy $\epsilon$. The square shows the
result evaluated from the DISTO Collaboration~\protect\cite{Balestra} 
while the triangles were obtained from the data 
of Ref.~\protect\cite{LB}. The dashed area  shows
the experimental data on the $pp{\to}pp\omega$ amplitude, divided 
by the factor 8.5, where the data are connected by a line through
their upper and lower error bars.}
\label{ozi2a}
\end{figure}

\begin{figure}[h]
\phantom{aa}\vspace{-0.3cm}
\psfig{file=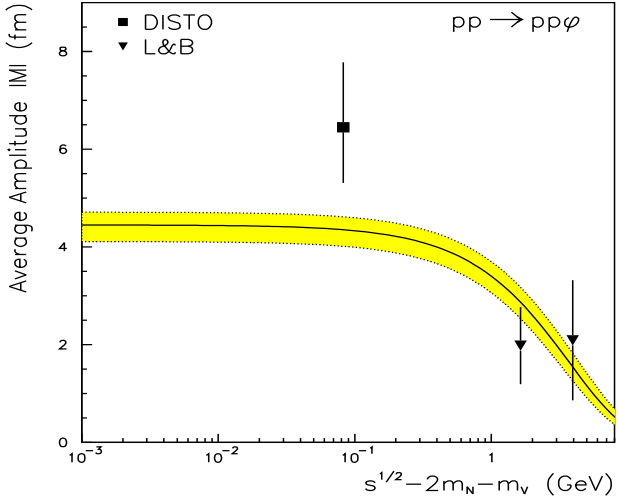,width=9.2cm,height=7.6cm}
\caption[]{The average amplitude $|M|$ for the $pp{\to}pp\phi$ reaction
as a function of the excess energy $\epsilon$. The solid line  shows the 
approximation~(\protect\ref{fit}) for the $pp{\to}pp\omega$ 
amplitude divided by the factor 8.5, while the dashed area is
the uncertainty of the approximation with respect to the 
$pp \rightarrow \omega pp$ data.}
\label{ozi2b}
\end{figure}

Now the DISTO data point~\cite{Balestra} for the  
$pp{\to}pp\phi$ total cross section
can be used for the evaluation of the reaction amplitude.
Fig. \ref{ozi2a} shows the experimental results for the average
$pp{\to}pp\phi$ production amplitude as a function of the excess
energy. Since there are only three experimental points  we 
cannot perform a statistical analysis of the 
$|M_\omega|$/$|M_\phi|$ ratio similar to the ${\pi}N \to VN$
analysis. Note that the $pp{\to}pp\phi$ data are available
only for $\epsilon{>}$80~MeV, where the FSI enhancement as 
well as the correction due to the final $\phi$-meson
width almost play no role.  

Now, to compare the data one might  take the
ratio of the $pp{\to}pp\omega$ and $pp{\to}pp\phi$
amplitudes as a constant. The two experimental points at
high energy give a  ratio $R \simeq$8.5. 
Fig.~\ref{ozi2a} shows the $pp \to pp\phi$ production 
amplitude together with the $pp \to pp\omega$ experimental 
results divided by the factor 8.5. To illustrate the
$\epsilon$-dependence the data are simply connected by upper 
and lower lines through their error bars. Fig.\ref{ozi2b}, 
furthermore, shows the 
data for the $pp{\to}pp\phi$ production amplitude using  
the fit~(\ref{fit}) for the $pp{\to}pp\omega$ amplitude again
divided by the factor 8.5. Here the DISTO data point sticks out from the
error band to some extent.
However, it is not clear if one might take the $\omega$/$\phi$
ratio as independent on $\epsilon$. As we already demonstrated for  the 
${\pi}N{\to}{\omega}N$ and ${\pi}N{\to}{\phi}N$ reactions,
the $|M_\omega|$/$|M_\phi|$ ratio substantially depends on
the excess energy for $\epsilon >$ 300 MeV. In this sense, 
the DISTO result does not strictly
contradict the $pp{\to}pp\phi$ data available at high energy.

\begin{figure}[h]
\phantom{bb}\vspace{-9mm}
\psfig{file=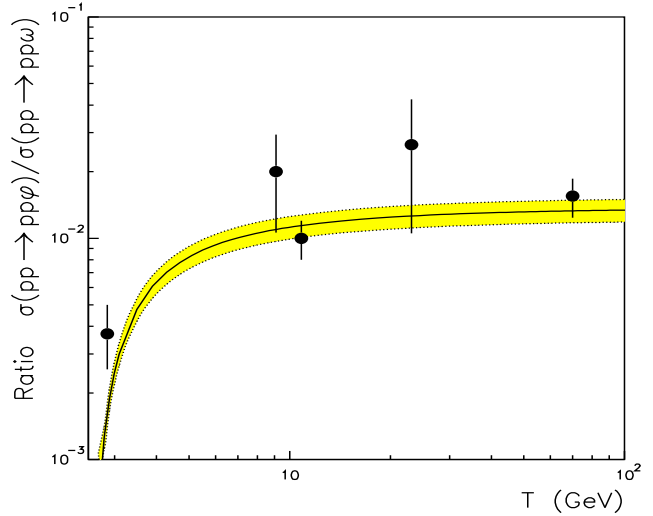,width=9.2cm,height=7.6cm}
\caption[]{The ratio of the $pp{\to}pp\phi$ and $pp{\to}pp\omega$
cross sections as a function of the beam energy $T$. Experimental data
are taken from Refs.~\protect\cite{LB,Balestra,Arenton,Golovkin}.
The solid line shows the result calculated with the energy independent 
ratio   $|M_\omega|$/$|M_\phi|$=8.3,
while the dashed area indicates the parent standard deviation.}
\label{ozi3}
\phantom{aa}\vspace{-4mm}
\end{figure}

Furthermore, since additional experimental 
results~\cite{Arenton,Golovkin} are available for the ratio of 
the $\phi$/$\omega$ total or differential cross sections above 8 GeV
bombarding energy, we also
show this ratio calculated with Eq.(\ref{sigma}) in Fig.\ref{ozi3} 
as a function of the incident proton energy. 

We have performed a $\chi^2$ fit to the available data on 
the ratio of the $pp{\to}pp\phi$ and $pp{\to}pp\omega$ cross sections
with a constant ratio of the $|M_\omega|$/$|M_\phi|$ production
amplitude and obtained the value of 8.5$\pm$1.0. Here the error
is due to the parent standard deviation. The confidence level of 
the fit is below 50\%. Again the DISTO result is not consistent 
with the constant ratio $|M_\omega|$/$|M_\phi|$=8.5.
We mention that the DISTO result on $\phi$-meson production can 
be fixed by 
$|M_\omega|$/$|M_\phi|$=5.72$^{+1.01}_{-1.17}$ with the 
$pp{\to}pp\omega$ amplitude taken from the approximation~(\ref{fit}).

\section{Theoretical interpretations}

\begin{figure}[h]
\phantom{aa}\vspace{-1.4cm}\hspace{-5mm}
\psfig{file=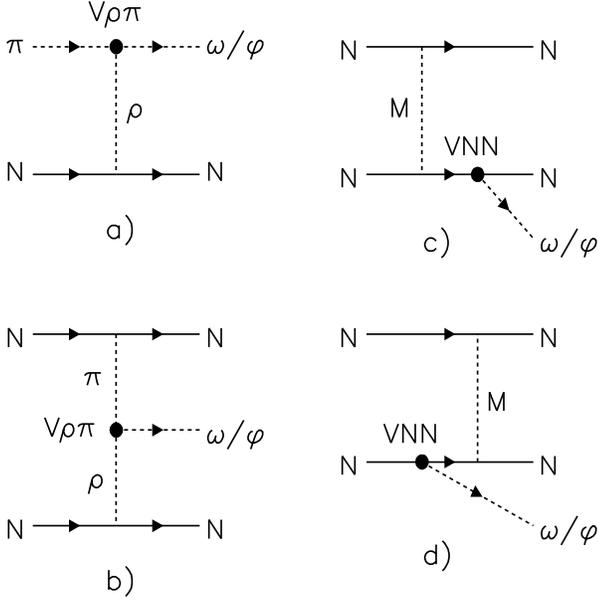,width=10cm,height=10.cm}
\phantom{aa}\vspace{-0.9cm}
\caption[]{The diagrams for the ${\pi}N{\to}VN$ (a) and 
$NN{\to}VNN$ (b-d) reactions with $V{=}\omega,\phi$, that contain 
the $V\rho\pi$ and $VNN$ vertices. Here $M$ denotes the $NN$
interaction in the initial or final state due to meson 
exchange.}
\label{ozi12}
\end{figure}

In general~\cite{Ellis} the experimental results on the $\phi{/}\omega$
ratio are compared to a constant as given by Lipkin~\cite{Lipkin},
\begin{eqnarray}
R^2(\phi / \omega) = \frac{g^2_{\phi\rho\pi}}{g^2_{\omega\rho\pi}}=
\frac{g^2_{\phi NN}}{g^2_{\omega NN}}=
\frac{\sigma(\pi N\to\phi X)}{\sigma(\pi N\to\omega X)} \nonumber \\
=\frac{\sigma(NN\to\phi X)}{\sigma(NN\to\omega X)}=tan^2 
(\Delta\theta_V)=4.2 \times 10^{-3},
\label{ozir1}
\end{eqnarray}
where $\Delta\theta_V{=}3.7^0$~\cite{PDG} is the deviation from the ideal 
$\omega{-}\phi$ mixing angle.
It is important to note, that Eq.~(\ref{ozir1}) provides 
the $\phi{/}\omega$ ratio  for hadronic reactions 
which can be expressed by the diagrams shown in Fig. \ref{ozi12}
that contain the $V\rho\pi$ and $VNN$ vertices. 

Furthermore, the ratio of the $\omega\rho\pi$ to $\phi\rho\pi$ coupling 
constant can be evaluated from the relevant partial decay 
width~\cite{Sakurai1,GellMann1}. The $\phi\rho\pi$ coupling constant 
can be measured (as  first proposed by Sakurai~\cite{Sakurai1}) by the 
$\phi{\to}\rho\pi$ decay via
\begin{eqnarray}
\Gamma_{\phi\to\rho\pi}&=&\frac{g_{\phi\rho\pi}^2}
{16\pi^2 m_\phi^5}\!\! \intop_{2m_\pi}^{m_\phi-m_\pi} \!\!d\mu \
\lambda^{3/2}(m_\phi^2,\mu^2,m_\pi^2) \nonumber \\
&\times&\frac{\mu^2 \ \Gamma_{\rho{\to}2\pi}(\mu)}
{(\mu^2-m_\rho^2)^2+\mu^2\Gamma_{\rho{\to}2\pi}^2(\mu)}. 
\label{phirhopi}
\end{eqnarray}
Taking into account the energy dependence of the $\rho$-meson width and
experimental numbers from the PDG~\cite{PDG} we obtain
$g_{\phi\rho\pi}$ as shown in Table~\ref{tab2}.

\begin{table}[h]
\caption{\label{tab2} The coupling constants and their sources of
extraction. The decay widths and  masses
are taken from Ref.~\protect\cite{PDG}. Taking into account the 
contribution from the $\omega{\to}3\pi$ decay, which is 
20\% at 90\% confidence level~\protect\cite{PDG}, we
obtain  $g_{\phi\rho\pi} \approx$ 1.1.}
\center{
\begin{tabular}{|c|l|c|}
\hline\noalign{\smallskip}
Vertex & Source & Constant  \\
\noalign{\smallskip}\hline\noalign{\smallskip}
$\phi \rho \pi $ & $\Gamma (\phi \to \rho \pi )$ & $1.23 \pm 0.05 $ \\
$\rho \gamma $ & $\Gamma (\rho \to e^+e^- )$ & $2.41\pm 0.12 $ \\
$\rho \gamma $ & $\Gamma (\rho \to \mu^+ \mu^-)$ & $ 2.45 \pm 0.15 $ \\
$\omega \gamma $ & $ \Gamma (\omega \to e^+e^- )$ & $ 8.24 \pm 0.24 $ \\
$\omega \gamma $ & $ \Gamma (\omega \to \mu^+ \mu^-)$ & $ >5.29 $ \\
$\omega \rho \pi $ & $\Gamma (\omega \to \pi^0 \gamma )$ & 
$8.82\pm 0.50 $ \\
$\omega \rho \pi $ & $\Gamma (\rho \to \pi^0 \gamma )$ &
$12.32 \pm 3.12 $ \\
$\omega \rho \pi $ & $\Gamma (\omega \to 3\pi) $ & $11.79 \pm 0.19 $ \\
\noalign{\smallskip}\hline
\end{tabular}}
\end{table}

The separate $\omega{\to}\rho\pi$ decay is not energetically 
allowed and  to determine the $\omega\rho\pi$ coupling 
constant Gell-Mann and Zachariasen~\cite{GellMann1} proposed 
to study the radiative decays
$\omega{\to}\pi\gamma$  and $\rho{\to}\pi\gamma $. In their approach 
(see also the review of Mei{\ss}ner~\cite{Meissner}) this
process is dominated by the $\omega\rho\pi $ vertex
with the intermediate vector meson  coupled to the
photon via  vector dominance.
The  $\omega\rho\pi $ coupling constant 
can be measured by~\cite{GellMann1,Kaymakcalan},
\begin{equation}
\label{lepton}
\Gamma (\omega \to \pi^0 \gamma )= 
\frac {g^2_{\omega \rho \pi}}{96 \ m_{\omega}^5}  
\ \frac{\alpha}{\gamma_{\rho}^2} \  
{\left\lbrack m_{\omega}^2-m_{\pi}^2 \right\rbrack}^3,
\end{equation}
where $\alpha$ is the fine structure constant. Furthermore, a
direct measurement of ${\gamma}_{\rho}$ 
is possible by means of the vector meson   
decay into leptons~\cite{Nambu}
\begin{equation}
\label{vdm1}
\Gamma (\rho \to l^+l^-)= \frac{\pi}{3} \
{\left\lbrack \frac {\alpha}{{\gamma}_{\rho}} \right\rbrack}^2 \ 
\sqrt{m_V^2-4m_l^2}
\left\lbrack 1+\frac{2m_l^2}{m_{\rho}^2} \ \right\rbrack ,
\end{equation}
where $m_{\rho}$ and $m_l$ are the masses of the vector meson and 
lepton, respectively. In a similar way  $g_{\omega\rho\pi}$
can be measured via the $\rho\to\pi^0 \gamma $ decay.
The relevant coupling constants obtained with the latest PDG 
fit to experimental data are listed in Table~\ref{tab2}.

On the other hand, Gell-Mann, Sharp and 
Wagner~\cite{GellMann2} proposed to determine 
$g_{\omega\rho\pi}$ through the $\omega\to3\pi $ decay 
assuming that the $\omega$ first
converts into $\rho\pi$ followed by $\rho\to 2\pi$.
The relation between the $\Gamma(\omega\to3\pi)$ and 
$\omega\rho\pi$ coupling constants is given in 
Ref.~\cite{Lichard}. A more elaborate analysis of the
$\omega{\to}3\pi $ decay includes  the  four-point contact 
term due to the direct coupling between the $\omega$-meson and 
three pions~\cite{Meissner,Klingl1,Kaymakcalan},
however, the contribution from this anomalous coupling
to $\Gamma(\omega\to3\pi)$ is only about 10\%. The analysis
from Refs. \cite{Klingl1,Klingl2} provides $g_{\omega\rho\pi}{=}10.88$.

Note that the mixing angle  can also be 
determined by the ratio of
the $\omega{\to}\pi^0\gamma$ and $\phi{\to}\pi^0\gamma$
radiative decay widths by applying vector 
dominance~(\ref{lepton}), which gives 
$g_{\omega\rho\pi}{/}g_{\phi\rho\pi}$= 12.9$\pm$0.4. An alternative
model~\cite{Meissner,Klingl1,Meissner2} proposed a 
direct $\omega\pi\gamma$  coupling, instead of the vector dominance,
where the ratio of $g_{\omega\rho\gamma}$ to 
$g_{\phi\rho\gamma}$ yields $16.8{\pm}1.0$. 
Both models predicts  values close to the mixing angle
$\theta_V{=}37^0$, determined from the mass splitting in the 
vector-meson nonet, but depend on the vector dominance or
direct coupling assumption. The direct $\phi{\to}\rho\pi$ decay
is a more standard way, although it leads to a rather
large uncertainty in the determination of the $\phi\rho\pi$
coupling. 

\begin{figure}[h]
\psfig{file=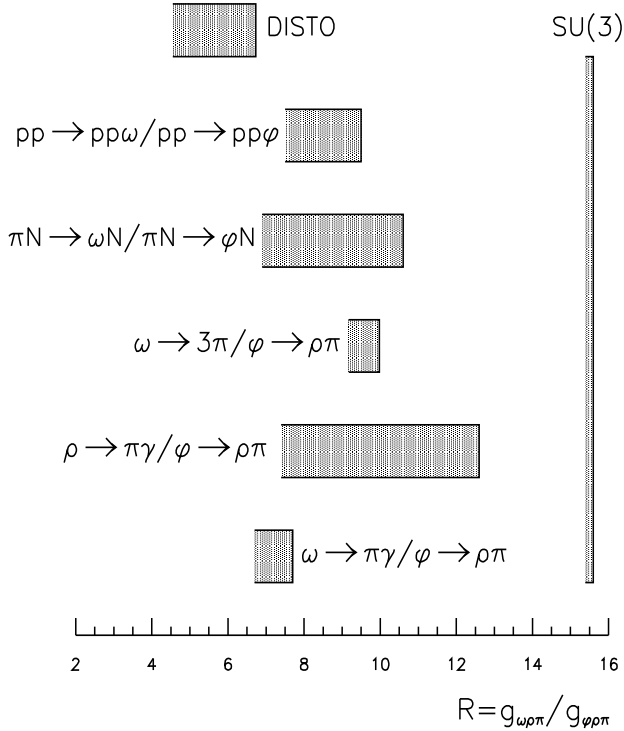,width=9.2cm,height=10cm}
\phantom{aa}\vspace{-0.1cm}
\caption[]{The ratio of the $\omega\rho\pi$ and $\phi\rho\pi$ coupling
constants evaluated from different sources of experimental data
in comparison to the SU(3) prediction for $\Theta_V = 39^0$.}
\label{ozi13}
\end{figure}

To provide a graphical overview, Fig.~\ref{ozi13} 
illustrates the ratio of the $\omega\rho\pi$ and 
$\phi\rho\pi$ coupling constants evaluated from the partial decay 
width. We also show the ratio given by the ${\pi}N{\to}VN$ and
$pp{\to}Vpp$ data assuming that this ratio is energy independent.
The DISTO result is shown separately and -- as discussed above -- 
is not consistent with the other data for $pp$ reactions. 
However, within the
present uncertainties the experimental results -- as evaluated from
all different sources -- appear to be compatible; they all disagree
with the SU(3) estimate based on the $\omega{-}\phi$ 
as given by the PDG~\cite{PDG}.

We note, furthermore, that  any  production 
mechanism different from those in Fig.~\ref{ozi12} will 
invalidate the overall scaling based on the $R^2(\phi{/}\omega)$ 
function~\cite{Nakayama2,Nakayama1}. For instance, as found in 
Refs.~\cite{Locher,Buzatu1,Mull,Buzatu2,Gortchakov,Anisovich,Markushin},
two-step processes with intermediate $K\bar{K}$, 
$K^\ast\bar{K}$  $K^\ast\bar{K^\ast}$ states
may contribute substantially  to $\phi$ production in
antiproton-proton annihilation. Certainly, such OZI allowed 
processes could have also an effect on $\phi$-meson production
in $\pi N$ and $NN$ reactions, but their actual contribution so far
is unknown here. In view of Fig. 3a we speculate that  
their contribution should be rather low for excess energies 
$\epsilon\leq$300 MeV.

\section{Summary}
We have analyzed the experimental data available for $\omega$ and
$\phi$-meson production from ${\pi}N$ and $pp$ reactions and have
evaluated the ratio of the reaction amplitudes. Indeed the
experimental $\phi$/$\omega$ ratio substantially deviates from the 
SU(3) estimate $R^2(\phi{/}\omega){=}4.2{\times}10^{-3}$, which 
is  based on the $\omega{-}\phi$ mixing angle of $\theta_V$=39$^0$. 

However, it is important to recall that this SU(3) estimate is given 
by the ratio of the $\phi\rho\pi$ to $\omega\rho\pi$ 
and $\phi NN$ to $\omega NN$ coupling
constants and is related only to the reaction mechanisms
involving the relevant $V\rho\pi$ and $VNN$ vertex. 
Obviously,  any other production 
mechanism~\cite{Locher,Buzatu1,Mull,Buzatu2,Gortchakov,Anisovich,Markushin} as well as  
different form factors in the $V\rho\pi$ and $VNN$
vertices will lead to a deviation of the experimental
ratios from the simple
scaling $R^2(\phi{/}\omega){=}4.2{\times}10^{-3}$.

On the other side, by fitting the experimental ratio
with a constant, 
our comparison of the ${\pi}N$ and $pp$ data
with the ratio of the $\phi\rho\pi$ and $\omega\rho\pi$ coupling
constant (as evaluated from the measured partial decay) shows
an overall compatibility. The full analysis indicates that 
-- within the experimental uncertainties -- the data on the 
partial decays 
as well as on ${\pi}N$ and $pp$ reactions  provide an average ratio
$R^2(\phi{/}\omega){\simeq}1.6{\times}10^{-2}$, which is close to
the DISTO data point, however, disagrees 
with the SU(3) estimate based on the $\omega{-}\phi$ mixing angle 
of $\theta_V$=39$^0$. 

\acknowledgement{
We appreciate valuable discussions with W. K\"uhn and J. Ritman
as well as comments and suggestions from C. Hanhart and 
J. Haidenbauer.}


\begin{thebibliography}{99}
\bibitem{Okubo} 
        S. Okubo, Phys. Lett.  \textbf{5}, (1963) 165.
\bibitem{Zweig} 
        G. Zweig, CERN report  TH-401 (1964).
\bibitem{Iizuka} 
        J. Iizuka, Prog. Theor. Phys. Suppl.  \textbf{38}, (1966) 21
\bibitem{PDG}
        Particle Data Group, Eur. Phys. J. C \textbf{3}, (1998) 1.
\bibitem{Lipkin}
        H.J. Lipkin, Phys. Lett. B \textbf{60},  (1976) 371. 
\bibitem{Balestra}
        F. Balestra et al., Phys. Rev. Lett. \textbf{81}, (1998) 4572.
\bibitem{LB}
        Landolt-B\"ornstein, \textit{New Series} \textbf{ I/12}
        (Springer, 1998).
\bibitem{Feynman}
        R.P  Feynman, \textit{Theory of Fundamental Processes}
        ( W.A. Benjamin Inc., New  York,  1962).
\bibitem{Hanhart}
        C. Hanhart, A. Kudryavtsev, nucl-th/9812022
\bibitem{Karami}
        H. Karami et al., Nucl. Phys. B \textbf{154}, (1979) 503.
\bibitem{Watson}
        K.M. Watson, Phys. Rev. \textbf{88}, (1952) 1163. 
\bibitem{Migdal}
        A.B. Migdal,  JETP \textbf{1}, (1955)  2.
\bibitem{GellMann}
        M. Gell-Mann and K.M. Watson, Ann. Rev. Nucl. Sci.
        \textbf{4}, (1954) 219. 
\bibitem{Taylor}
        J.R. Taylor, \textit{Scattering Theory} (Willey, New York, 1972).
\bibitem{Hibou}
        F. Hibou et al., nucl-ex/9903003.
\bibitem{DKT}
        B.L. Druzhinin, A. Kudryavtsev, and V.E. Tarasov, Z. Phys.,
        A \textbf{359}, (1997) 205.
\bibitem{SC1}
        A. Sibirtsev and W. Cassing, nucl-th/9904046.
\bibitem{SC2}
        A. Sibirtsev and W. Cassing, Eur. Phys. J. A \textbf{2}, 
        (1998) 333.
\bibitem{Arenton}
        M.W. Arenton et al.,  Phys. Rev. D \textbf{25}, (1982) 2241.
\bibitem{Golovkin}
        S.V. Golovkin et al., Z. Phys. A \textbf{359}, (1997) 435.
\bibitem{Ellis}
        J. Ellis et al., Phys. Lett. B \textbf{353}, (1995) 319.
\bibitem{Sakurai1}
        J.J. Sakurai, Phys. Rev. Lett. \textbf{9}, (1962) 472.
\bibitem{GellMann1}
        M. Gell-Mann and F. Zachariasen, Phys. Rev. \textbf{124},
        (1961) 953.
\bibitem{Meissner}
        Ulf-G. Mei{\ss}ner, Phys. Rep. \textbf{161}, (1988) 213.
\bibitem{Kaymakcalan}
        \"O. Kaymakcalan, S. Rajeev and J. Schechter,
        Phys. Rev. D \textbf{30}, (1984) 594.
\bibitem{Nambu}
        Y. Nambu and J.J. Sakurai, Phys. Rev. Lett. \textbf{8}, (1962) 79.
\bibitem{GellMann2}
        M. Gell-Mann, D. Sharp and W.G. Wagner, Phys. Rev. Lett.
        \textbf{8}, (1962) 261.
\bibitem{Lichard}
        P. Lichard, Phys. Rev. D \textbf{49}, (1994) 5812.
\bibitem{Klingl1}
        F. Klingl, N. Kaiser and W. Weise, Z. Phys. A \textbf{356}, 
        (1996) 193.
\bibitem{Klingl2}
        F. Klingl, private communication.
\bibitem{Meissner2}
        P. Jain et al., Phys. Rev. D \textbf{37}, (1998) 3252.
\bibitem{Nakayama2}
        K. Nakayama et al., nucl-th/9904040.       
\bibitem{Nakayama1}
        K. Nakayama et al., Phys. Rev. C \textbf{57}, (1998) 1580.
\bibitem{Locher}
        M.P. Locher, Y. Lu and B.S. Zou, Z. Phys. A \textbf{347},
        (1994) 281.
\bibitem{Buzatu1}
        D. Buzatu and F.M. Lev, Phys. Lett. B \textbf{329}, (1994)
        143.
\bibitem{Mull}
        V. Mull, K. Holinde and J. Speth, Phys. Lett. B \textbf{334},
        (1994) 295.
\bibitem{Buzatu2}
        D. Buzatu and F.M. Lev, Phys. Rev. C \textbf{51}, (1995)
        2893.
\bibitem{Gortchakov}
        O. Gortchakov, M.P. Locher, V.E. Markushin and S. von Rotz,
        Z. Phys. A \textbf{353}, (1996) 447.
\bibitem{Anisovich}
        A.V. Anisovich and E. Klempt, Z. Phys. A \textbf{354},
        (1996) 197.
\bibitem{Markushin}
        V.E. Markushin and M.P. Locher, Eur. Phys. J. A \textbf{1},
        (1998) 91.
\end{thebibliography}
\end{document}